\documentclass[journal]{IEEEtran}

\ifCLASSINFOpdf
\else
\fi
\usepackage{graphicx}
\usepackage{grffile}
\usepackage{epstopdf}
\usepackage{amsmath}
\usepackage{amssymb}
\usepackage{amsthm}
\usepackage{subfig}
\usepackage{multirow}
\usepackage[font=footnotesize,labelfont=]{caption}

\begin{document}

\title{An Investigation of Randomized Controlled Trial (RCT) Method as a Customer Baseline Load (CBL) Calculation for Residential Customers}

\author{Saeed Mohajeryami
	\thanks{Saeed Mohajeryami 
	is with the Department of Electrical and Computer Engineering, University of North Carolina at Charlotte, Charlotte, NC 28223 USA e-mail: smohajer@uncc.edu
	

}
	\thanks{}
	\thanks{}
	}

\markboth{}%
{Shell \MakeLowercase{\textit{et al.}}: Bare Demo of IEEEtran.cls for IEEE Journals}

\maketitle

\begin{abstract}
FERC Order 745 allows demand response owners to sell their load reduction in the wholesale market. However, in order to be able to sell the load reduction, some implementation challenges must be addressed, one of which is to establish Customer Baseline Load (CBL) calculation methods with acceptable error performance, which has proven to be very challenging so far. In this paper, the error and financial performance of Randomized Controlled Trial (RCT) method, applied to both granular and aggregated forms of the consumption load, are investigated for a hypothetical demand response program offered to a real dataset of residential customers .   
   
\end{abstract}

\IEEEpeerreviewmaketitle

\section{Introduction}
\IEEEPARstart{I}{n} response to the inefficient functionality of the wholesale electricity  market in the absence of demand participation, Federal Energy Regulatory Commission (FERC) has strongly encouraged the adoption of Demand Response (DR) programs in FERC No. 745 order \cite{FERC745}. These programs are designed to incentivize customers to temporarily reduce their demand in answer to price signals. DR programs, according to numerous studies, can provide workable solutions to many major problems in the power system. Stabilizing the wholesale prices, limiting the market power of large players, ensuring the reliability during emergencies, and providing balancing act to address the variability of renewable energies are a few notable examples of DR programs' benefits. \cite{aghaei,Moh1}. 

Although these programs, in theory, appear to be fairly simple and straightforward, there are many obstacles that pose challenges to the implementation in practice, chief among them Evaluation Measurement and Verification (EM\&V) of the customers' load reduction. In order to perform the payment settlement, which is the critical part of all DR programs, it is essential that the load reduction is accurately measured and verified. To carry out such task, Customer Baseline Load (CBL) is to reliably be estimated. The CBL is the amount of electricity that customers would have consumed in the absence of the DR event day. If the CBL is calculated accurately, then the real load reduction could be measured as the difference between actual consumption and the CBL. 

The accurate calculation of CBL is extensively investigated in the literature. In addition, FERC 745 order has requested Independent System Operation (ISOs) to establish a CBL calculation method for all their DR participants \cite{FERC745}. Some of the well-established methods employed by different ISOs are summarized in Table \ref{CBLmethods}. These methods are fully described in \cite{Moh2}. It is necessary to mention that these CBL calculation methods are originally developed for large industrial and commercial customers. 

\begin{table}[b!]
\centering
\caption{CALCULATION METHODS EMPLOYED BY DIFFERENT ISOs}
\label{CBLmethods}
\begin{tabular}{|c|c|}
\hline
\textbf{Independent System Operator (ISO)} & \textbf{CBL Calculation Method} \\ \hline
PJM & Averaging (High4of5) \\ \hline
NYISO & Averaging (High5of10) \\ \hline
CAISO & Averaging (High10of10) \\ \hline
Ontario,Canada & Averaging (High15of20) \\ \hline
ISONE & Exponential Moving Average \\ \hline
ERCOT & Regression Models \\ \hline
\end{tabular}
\end{table}

In recent years, high penetration of smart meters in residential sector, which provide granular hourly consumption data, has created unprecedented opportunities for load aggregators to enter into contract with residential customers and to offer their load reduction as a supply source in the wholesale market. 
In contrast to ISOs' DR programs that work with large industrial and commercial customers, the load aggregators mainly deal with residential customers.
Developing CBL calculation methods for residential customers faces more challenges as the load curve of these customers have much more random characteristics compared to large industrial and commercial customers. This randomness is driven by the multitudes of non-correlated personal and household activities. By taking the fluctuations that exist in residential customers' data into consideration, specific CBL calculation methods ought to be improved in a way that the effect of such volatility is addressed. The authors in \cite{Moh3,Moh4,Moh5,wijaya} show that CBL methods that are developed for large industrial and commercial customers appear to make an unsatisfactory error for residential customers. To date, developing CBL methods for residential customers has rarely been seriously scrutinized in the literature. 

The authors in \cite{sharifi}, have carried out a minor modification on existing CBL calculation methods to make them suitable for residential customers. They propose that existing CBL methods should be adjusted according to the hourly temperature on the event day. Nevertheless, the authors fail to verify their proposed method with any validation experiments. 

In addition to the methods demonstrated in Table \ref{CBLmethods}, Randomized Controlled Trial (RCT) has been employed as a tool for the CBL calculation. This method, unlike the other available CBL calculation methods, is observed to only be utilized for residential customers. This method is recommended by Lawrence Berkeley National Lab (LBNL) as one of the methods that could be used to assess the effects of time-based rates, enabling technologies, and various other treatments on customers' consumption levels and patterns of usage \cite{cappers}. Furthermore, the authors in \cite{Todd}, have recommended using RCT for evaluating the energy efficiency (i.e. load reduction) in behavior-based efficiency programs. They assert that RCT method in comparison with the alternative methods is more robust and unbiased. Moreover, they acknowledge that due to the counterfactual nature of the real load reduction, it is impossible to measure it, and it can only be estimated; therefore, the RCT estimates would contain inherent randomness. The authors, however, do not provide any further discussions how RCT deals with this source of uncertainty (i.e. inherent randomness).

Moreover, in another attempt to use the RCT for EM\&V purposes, Green Mountain Power (GMP) electric utility company has employed the RCT for their consumer behavior study \cite{Blumsack}. GMP attempted to examine the impact of two pricing structures, Critical Peak Pricing (CPP) and Peak Time Rebate (PTR) on customers' consumption. These two programs were offered to residential customers during years 2012 and 2013. In CPP program, customers are charged extra amount of money for the consumption above their CBL, whereas in PTR, customers are rewarded monetarily for the load reduction (from their CBL level). In their study, the RCT is treated as a 100\% accurate method. According to the results, customers involved in the CPP program, reduced their average hourly loads by 5 to 15 \%, and customers who participated in PTR program reduced their average hourly loads by 5 to 8 \% \cite{Blumsack}. In addition, there are some other pilot projects that have used RCT as an EM\&V tool. During summer 2015, PG\&E, in partnership with Opower, conducted a Behavioral Demand Response (BDR) analysis to explore how customers could be engaged through communication and social comparison to reduce their peak load demand. By using RCT and aggregated loads, they found that on average this program can produce a 2.4\% reduction in peak usage \cite{Cook}. 

The main criticism of whether or not the RCT is accurate enough to capture the load changes in the aforementioned ranges remains unresponded in all aforesaid works. If the accuracy of the RCT is not carefully explored, the results achieved in these works could be entirely misleading. The gravity of this issue, which is largely neglected, is examined in this work. In fact, although the RCT is regarded by many as the gold standard for statistical testing of effects \cite{cappers}, it is necessary to investigate its performance for residential customers' load curves, either in granular or aggregated form, without any presupposition. In this paper, granular form refers to the programs that treat each customer individually and calculate an individual CBL for the purpose of payment settlement. On the other hand, aggregated form refers to the programs that do not engage with each customer in the individual level. These programs, first, aggregate the historical consumption data of a group of customers, then they make the CBL calculation and perform payment settlement in the aggregated level. 

In this paper, one of the goals of this paper is to assess the performance of the RCT method employed for residential customers, for both granular and aggregated load forms, and to compare the results with one of the well-established averaging CBL calculation methods (NYISO method, i.e. High5of10). Averaging methods are the most popular group of CBL calculation methods and are employed almost by all the ISOs. 

For this purpose, both error and financial performances of the methods are analyzed for a case of hypothetical PTR program offered to a real dataset of residential customers.  The details of the data will be elaborated in the following sections. PTR program is chosen because it depends on calculating CBL for the evaluation of load reduction \cite{Behavior}.  

The rest of the paper is organized as follows. An overview of CBL calculation methods is provided in section II. Afterwards, in section III, first, the dataset utilized for CBL calculation is introduced; then, three error metrics of accuracy, bias, and Overall Performance Index (OPI) are introduced and finally the experiment's setup is described. In section IV, the error analysis for two cases of  granular and aggregated load forms are presented. In section V, a case study for an economic analysis of the hypothetical PTR program is introduced. The results and discussion of the case study for both granular and aggregated cases are presented in section VI. The paper then concludes in section VII. 
  
\section{CBL calculation methods}
\begin{figure}[b!]
	\centering
	\includegraphics[clip,width=\columnwidth]{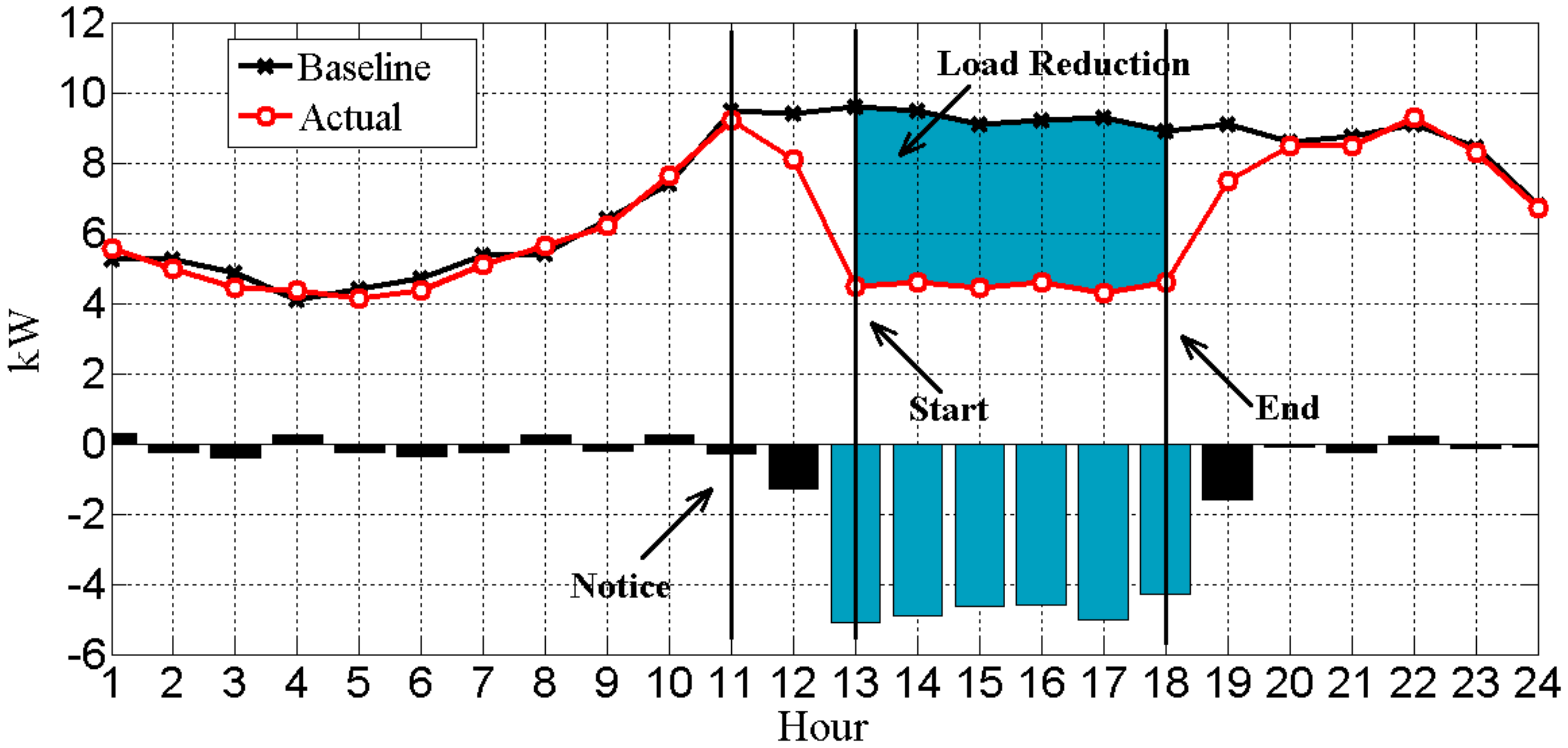}%
	\caption{Illustration of CBL, actual load and estimated load reduction}
	\label{CBL}
\end{figure}

In this section, two methods of HighXofY and RCT are briefly reviewed as tools for CBL calculation. As discussed earlier, CBL is the amount of load that is estimated to be consumed by customers in the absence of a DR curtailment signal. Fig. \ref{CBL} is the illustration of three concepts of actual load, CBL, and the estimated load reduction on the event day. In this figure, the black line represents the estimated baseline and the red line represents the actual consumption. The difference of these two line during the event hours is the amount of the load reduction, which is shaded by blue color. The difference of the black and red lines are shown by bar charts. Event days refer to the days that the ISO or RTO asks for DR bids. 

\subsection{HighXofY Method}
In this method, Y days of non-event, non-holiday weekdays and weekends prior to a DR-event day are selected. Then, X days with maximum average consumption are selected out of the Y days. The baseline is defined for each hour of the event day as the average hourly load of these X days. The New York ISO uses this method with X=5 and Y=10. In this paper, this method is selected with one modification. Since residential customers' consumption are not sensitive to weekends, the weekends, also, are included in the process of the CBL calculation. The algorithm of NYISO is described in \cite{grimm2008}.
\subsection{Randomized Controlled Trial (RCT)}
RCTs are very popular and trustworthy as evaluation methods to the extent that, as mentioned earlier, many regard them as the “gold standard” of evaluation methods. RCTs eliminate the selection effects by randomly assigning the households into two groups of treatment and control.   Since control and treatment groups are exposed to the similar conditions, the other possible alternative explanations will be eliminated, and the difference between these two groups could be attributed solely to the treatment. This way, the internal validity would be assured by the construction of the design \cite{EPRI}. 

The process starts by random assignment of households of the dataset into two groups. One group would serve as a basis for the calculation of CBL for the other group. It is essential that the customers do not exert any control over the assignment process (i.e. conscription) in order to ensure the internal and external validity of the results \cite{cappers}. 

RCTs could be enhanced by some procedures; however, the enhancement are prone to introduction of some factors affecting adversely the performance of such forms of RCTs. RCTs rely on minimal assumptions about the nature of customers; therefore, they can produce unbiased estimates of treatment effects. On the other hand, if these methods are enhanced with matching techniques like propensity score matching, nearest neighbor matching, etc., which require relying on some strong assumptions about the nature of customers, in case of the violation of these assumptions, RCT could produce biased results \cite{cappers}. 

The RCT method has lower administrative cost compared to NYISO as it requires no historical data for CBL calculation. Therefore, in the equal condition, RCT is much better alternative both in terms of lower cost and lower complexity. 
\section{Implementation}
In this section, the dataset employed for the error analysis is introduced and the error metrics are defined. Afterwards, these metrics are applied to CBL calculations and the results are presented.  
\subsection{Dataset}
In this paper, a dataset collected by Australian Energy Market Operation (AEMO) for 199 residential customers, in the leap year of 2012 (366 days), has been employed. Each electricity distributor in AEMO market supplies raw data from a sample of 200 customers in each of their supply areas to market operator in order to construct load profiles \cite{AEMO}. The data used in this paper is a sample of the raw data for one of the distributors. The customers under study are charged based on fixed tariff. The data used in this study is broken down into four seasons. Seasons in Australia are as follows: 
\begin{figure}[t!]
	\centering
	\includegraphics[clip,width=\columnwidth]{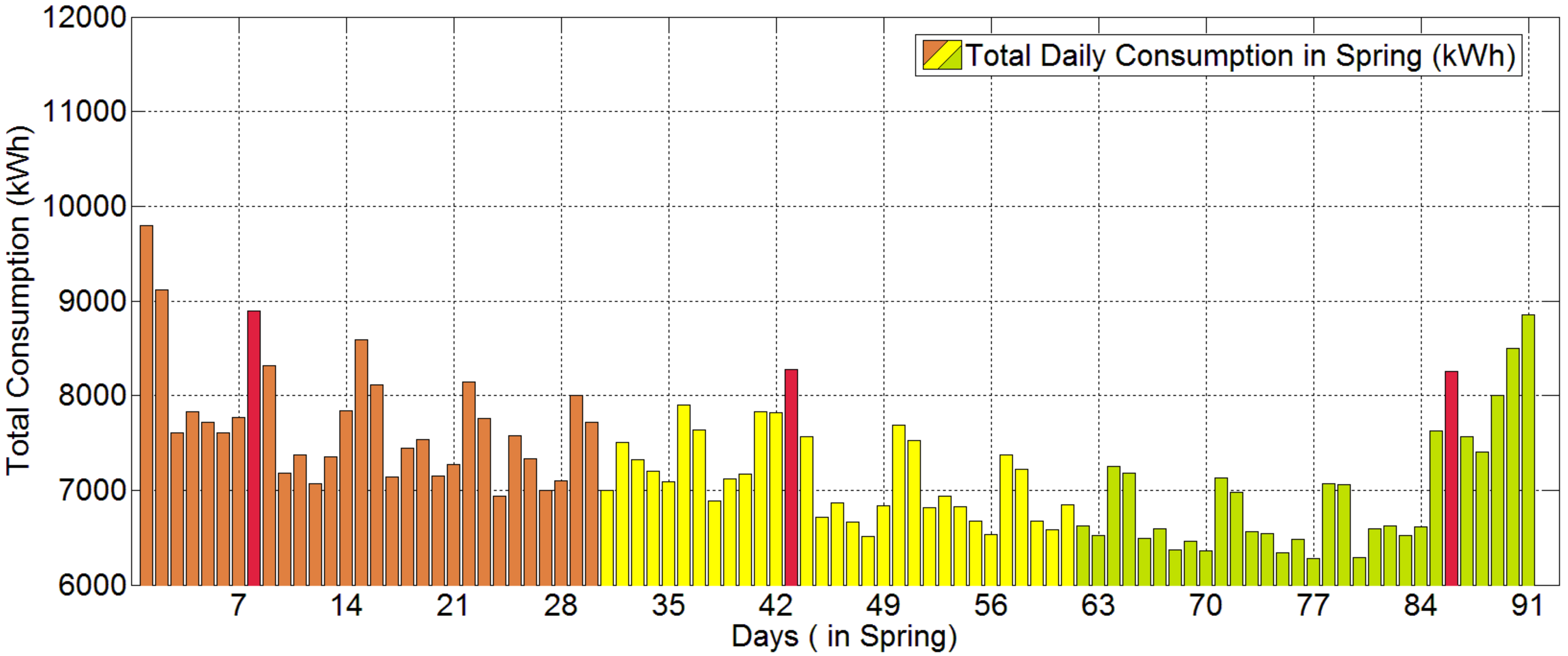}%
	\caption{Total daily consumption for the 199 customers in spring 2012(Orange for Sept., Yellow for Oct., light green for Nov.)}
	\label{spring}
\end{figure}

\begin{figure}[t!]
	\centering
	\includegraphics[clip,width=\columnwidth]{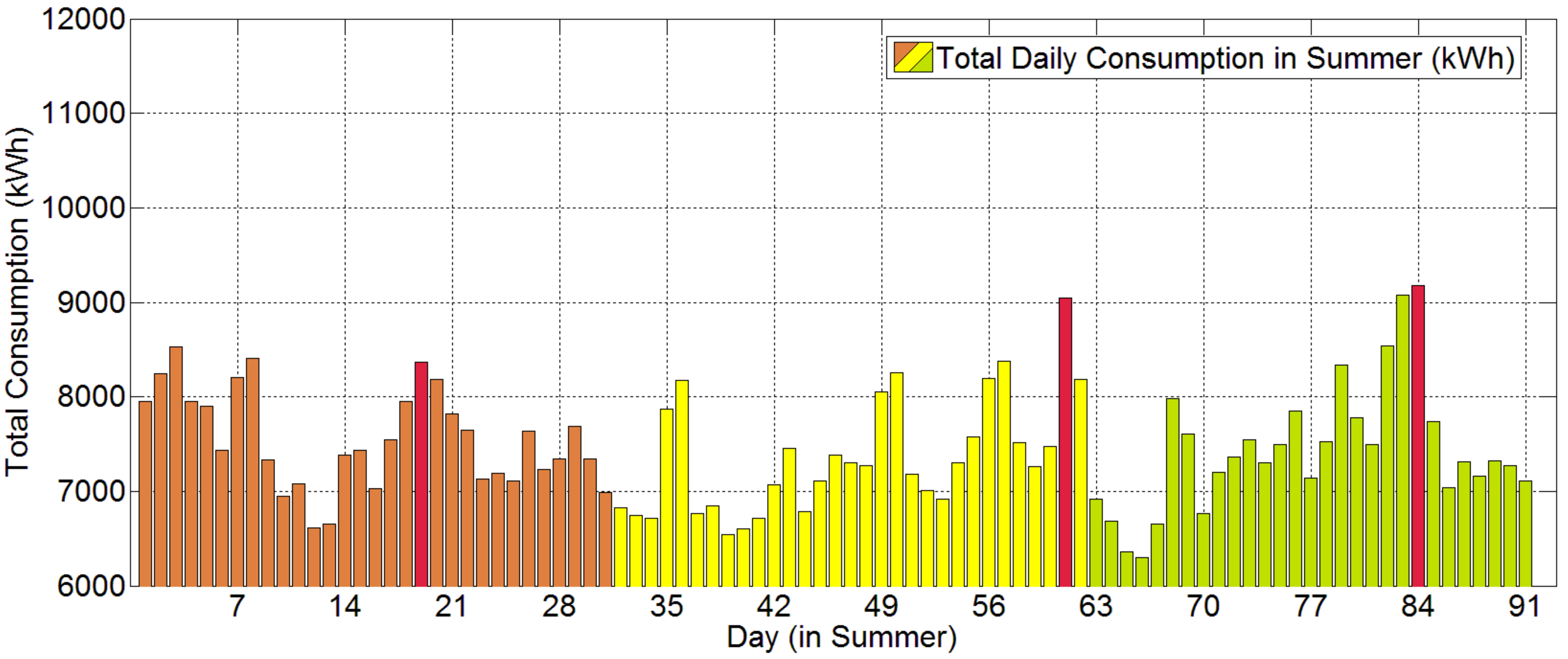}%
	\caption{Total daily consumption for the 199 customers in summer 2012(Orange for Dec., Yellow for Jan., light green for Feb.)}
	\label{summer}
\end{figure}

\begin{figure}[t!]
	\centering
	\includegraphics[clip,width=\columnwidth]{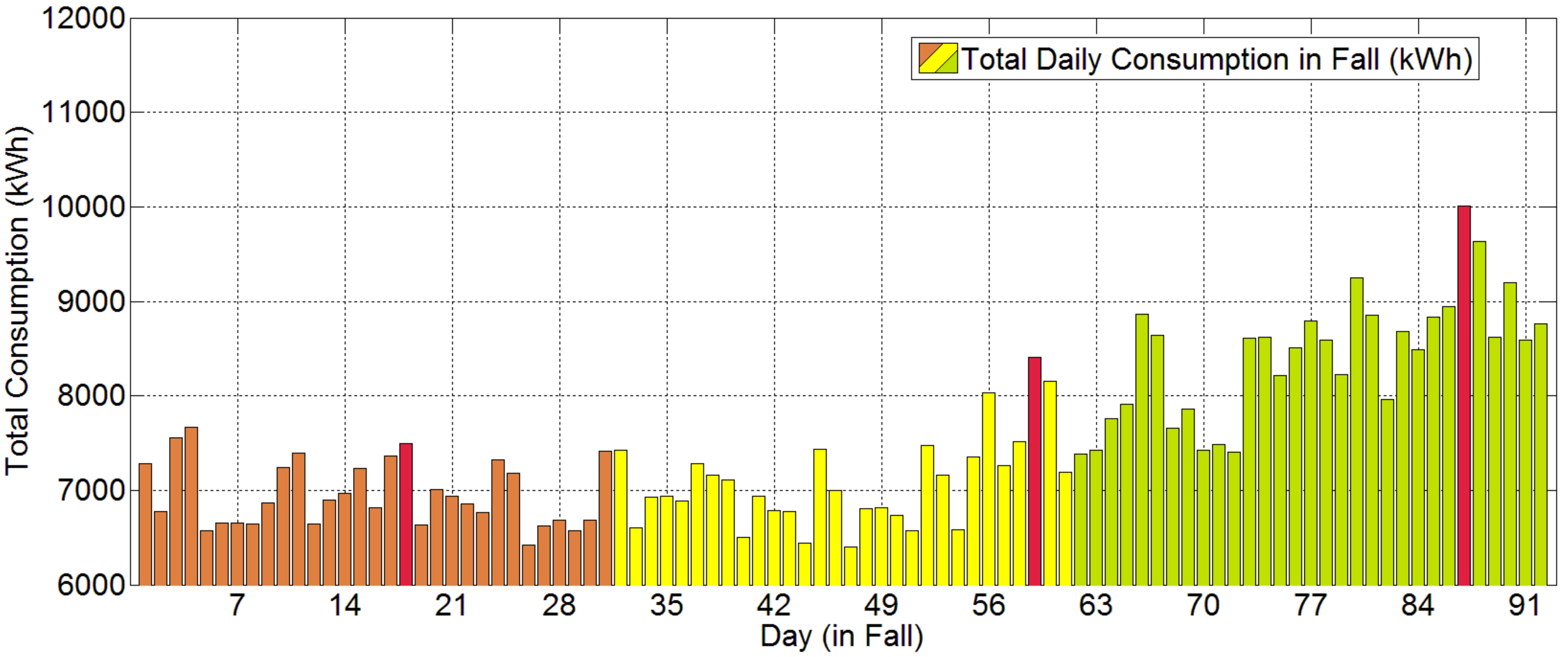}%
	\caption{Total daily consumption for the 199 customers in fall 2012(Orange for Mar., Yellow for Apr., light green for May.)}
	\label{fall}
\end{figure}

\begin{figure}[t!]
	\centering
	\includegraphics[clip,width=\columnwidth]{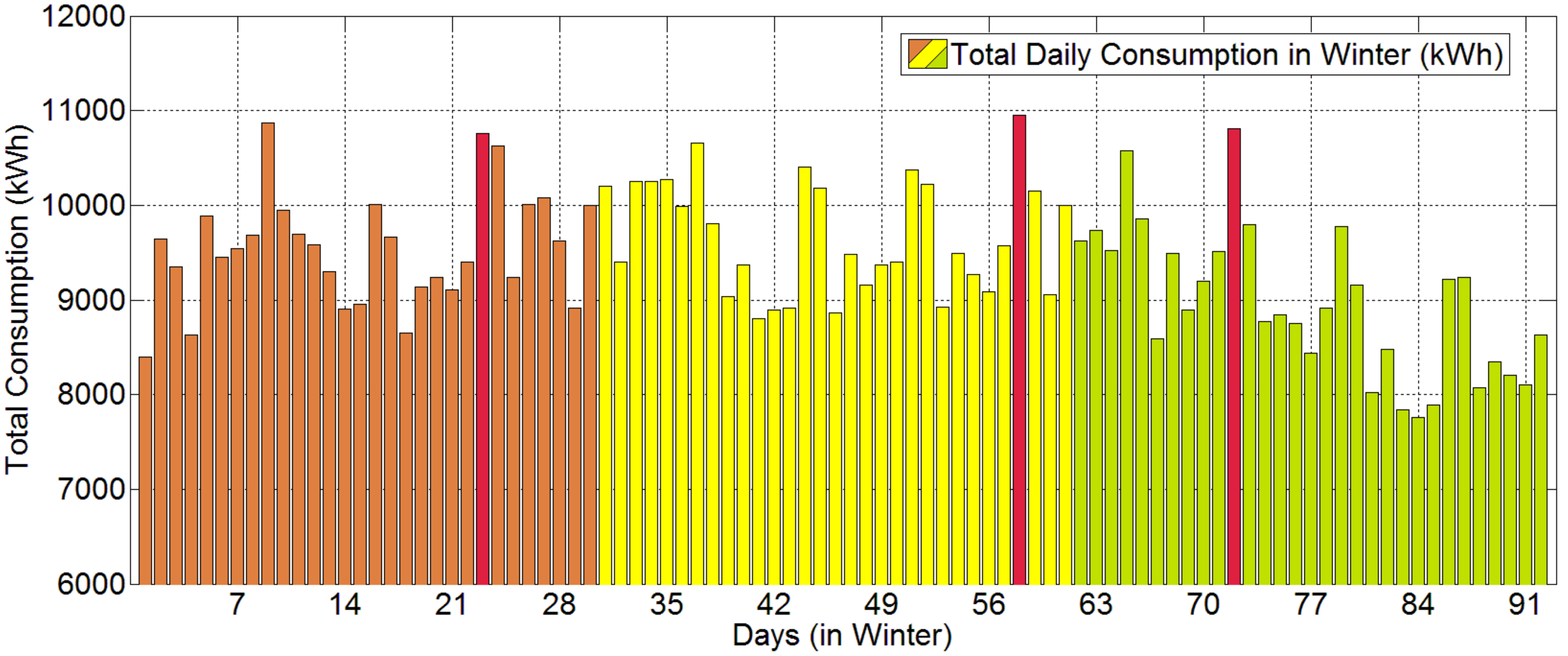}%
	\caption{Total daily consumption for the 199 customers in winter 2012(Orange for Jun., Yellow for Jul., light green for Aug.)}
	\label{winter}
\end{figure}

\begin{itemize}
	\item Spring - the three transition months September, October, and November.
\item Summer - the three hottest months December, January, and February.
\item Fall - the transition months March, April, and May.
\item Winter - the three coldest months June, July, and August.	
\end{itemize}

In this paper, 12 event days (one for each month) are selected for the error analysis, and the information about these days are as follows:
\begin{enumerate}
	\item Event day = Sept. 8th  (252nd  day)
\item Event day = Oct. 13th  (287th  day)
\item Event day = Nov. 25th  (330th  day)
\item Event day = Dec. 19th  (354th  day)
\item Event day = Jan. 30th  (30th  day)
\item Event day = Feb. 22nd  (53rd day)
\item Event day = Mar. 18th  (78th  day)
\item Event day = Apr. 28th  (119th  day)
\item Event day = May 25th  (147th  day)
\item Event day = Jun. 23rd  (175th  day)
\item Event day = Jul. 28th  (210th  day)
\item Event day = Aug. 11th  (224th  day)
\end{enumerate}

The total consumption for all four seasons are depicted in Figs. \ref{spring}-\ref{winter}. Event days are color-coded with red. The event hours are assumed to start from 3:00 p.m. and end at 9:00 p.m. Table \ref{averageconsumption}, shows the average consumption per capita for 12 event days in each treatment group. This information is useful to see the relationship between the inaccuracy and the average consumption per capita.

\begin{table}[b]
	\centering
	\caption{THE AVERAGE CONSUMPTION (kWh/hour) PER CAPITA IN EACH TREATMENT GROUP}
	\label{averageconsumption}
	\begin{tabular}{|c|c|}
		\hline
		\begin{tabular}[c]{@{}c@{}}Percent of \\ Treatment Population\end{tabular} & \begin{tabular}[c]{@{}c@{}}Average Consumption\\ (kWh/hour)\end{tabular} \\ \hline
		95\% & 1.84 \\ \hline
		90\% & 1.86 \\ \hline
		85\% & 1.90 \\ \hline
		80\% & 1.92 \\ \hline
		75\% & 1.96 \\ \hline
	\end{tabular}
\end{table}

\subsection{Error Metrics}
In this subsection, three error metrics of accuracy, bias, and OPI are introduced and elaborated. 

\subsubsection{Accuracy}
The hourly accuracy represents the hourly difference between the estimation and the real consumption. Let C be the set of all 199 customers, D be the set of all days in the data set, and T be the set of hourly timeslots in a day. Mean absolute error (MAE) for measuring baseline accuracy is defined as shown in (1). As shown, the lower the MAE, the higher the accuracy. 
\begin{equation}
\alpha =\frac{\sum_{i \in C}\sum_{d \in D}\sum_{t \in T} \left | b_{i}\left ( d,t \right )-l_{i}\left ( d,t \right ) \right  |}{\left | C \right |.\left | D \right |.\left | T \right |}
\label{eq7}
\end{equation}

\subsubsection{Bias}
Baseline bias is defined as shown in (2). The definition of bias is close to accuracy; however, it gives different information about the performance of CBL.
\begin{equation}
\beta=\frac{\sum_{i \in C}\sum_{d \in D}\sum_{t \in T} \left ( b_{i}\left ( d,t \right )-l_{i}\left ( d,t \right ) \right  )}{\left | C \right |.\left | D \right |.\left | T \right |}
\label{eq8}
\end{equation}

The difference between accuracy and bias, as expressed in (1) and (2) is the value of the difference between CBL and the actual consumption, which MAE uses the absolute value, while bias uses the real value. According to (2), baseline methods with positive bias overestimate the customers' actual consumption and vice versa. 

\subsubsection{Overall Performance Index (OPI)}
The overall error performance of a method depends on both accuracy and bias. Therefore, in this paper, another metrics is defined for measuring the overall performance. It is defined as the weighted sum of the absolute value of accuracy and bias, and it is called Overall Performance Index (OPI) as shown in equation (3). Moreover, the weight of  is selected for both absolute value of accuracy and bias, which indicates the equal importance of both accuracy and bias in the overall error performance.
\begin{equation}
OPI=\lambda  \left | \alpha  \right | + \left ( 1-\lambda  \right )\left | \beta  \right |
\label{eq9}
\end{equation}

\subsection{Setup}
One major concern in RCT is how to construct the control population. In this paper, in order to study how much the construct of control population impacts the error of CBL calculation methods, five separate datasets are created out of the original dataset. Each group has different number of customers in its control and treatment groups. The percent of control group in these five groups are 5\% (10 customers), 10\% (20 customers), 15\% (30 customers), 20\% (40 customers), and 25\% (50 customers). The treatment groups are consisted of the remaining customers. In the next subsection, the error performance of CBLs calculated for the treatment groups will be analyzed.

\section{Error Analysis}
In this subsection, the error performance of two CBL calculation methods for two cases of granular and aggregated loads are assessed.
\subsection{Granular Case}
\begin{figure}[h!]
	\centering
	\includegraphics[clip,width=\columnwidth]{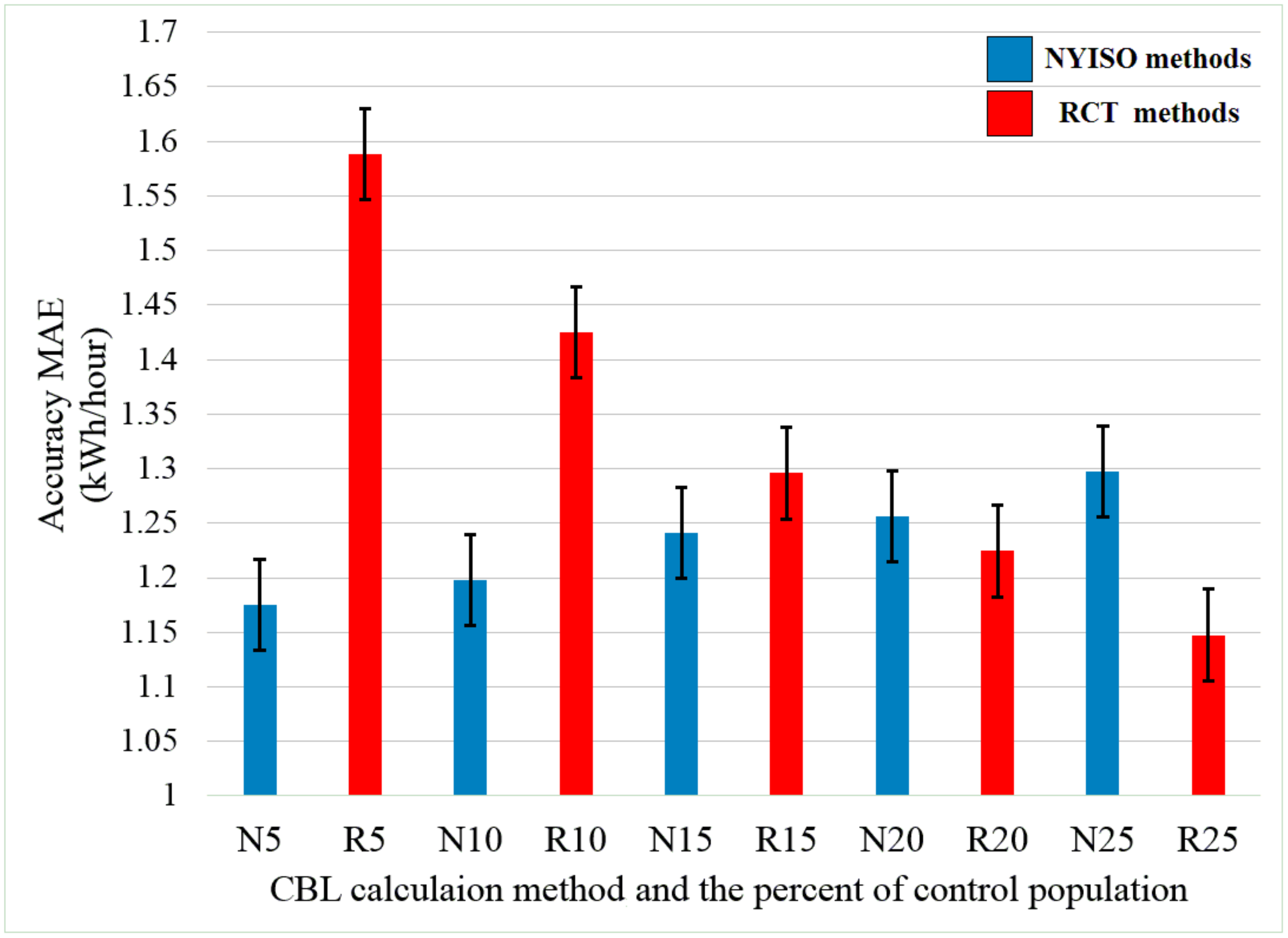}%
	\caption{Accuracy MAE for different CBL calculation with different percent of control population for granular case}
	\label{granularacc}
\end{figure}

\begin{figure}[t!]
	\centering
	\includegraphics[clip,width=\columnwidth]{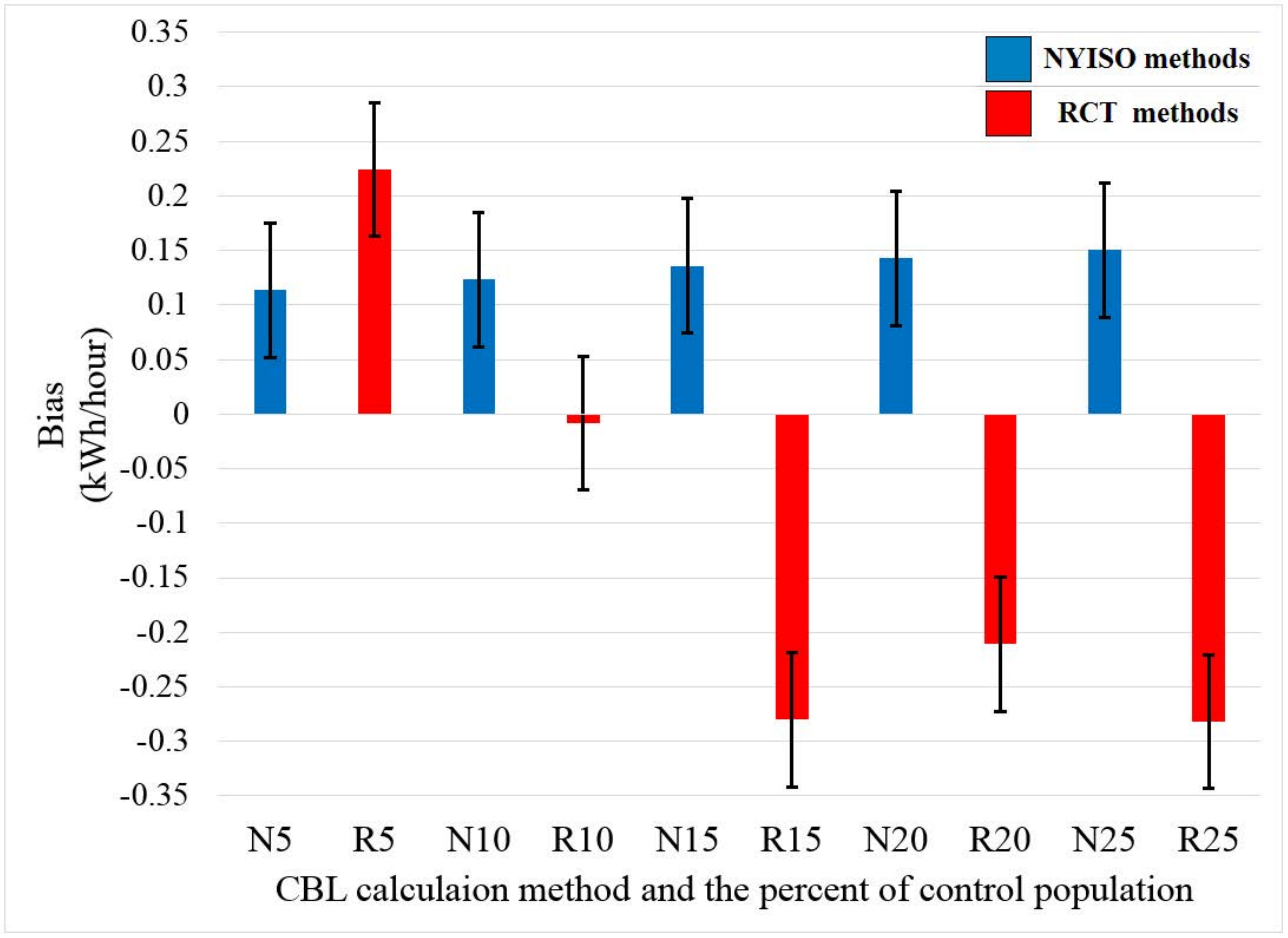}%
	\caption{Bias value for different CBL calculation with different percent of control population for granular case}
	\label{granularbias}
\end{figure}

\begin{figure}[t!]
	\centering
	\includegraphics[clip,width=\columnwidth]{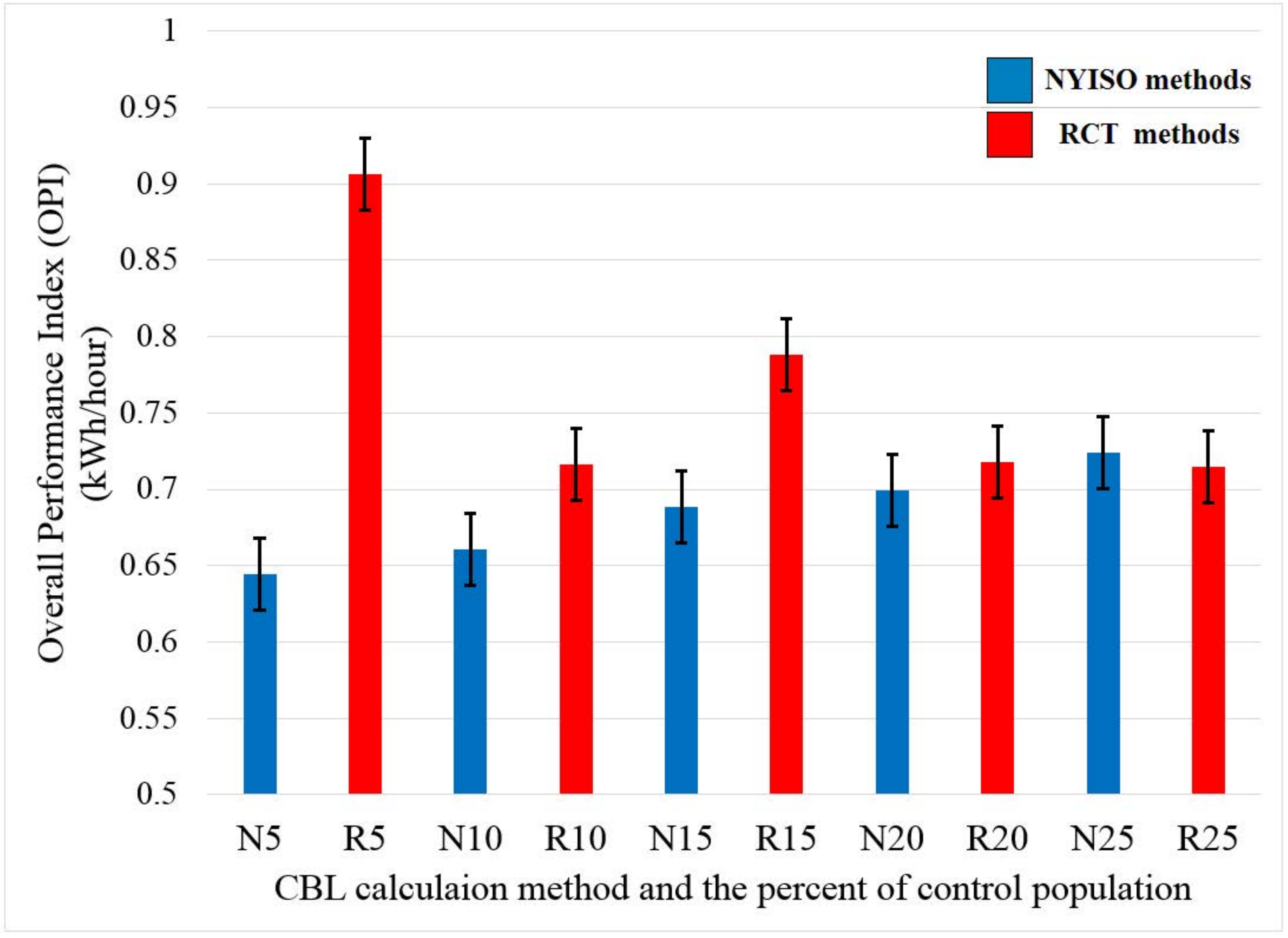}%
	\caption{OPI for different CBL calculation with different percent of control population for granular case}
	\label{granularOPI}
\end{figure}

As mentioned earlier, in the granular case, the PTR program calculates an individual CBL for each customer for the purpose individual payment settlement. Therefore, in this case, for all 12 event days, the CBL for all customers in different datasets is calculated. Table \ref{errorgranular} lists the value of three error metrics of the CBL calculations. The calculated values are for event hours. The information in this Table is, also, illustrated in three separate Figs. \ref{granularacc}-\ref{granularOPI}. Besides, whiskers (i.e. error bars) is included to each bar plot, which represents 95\% confidence interval around the adjusted mean for all 12 event days. According to the results in Fig. \ref{granularacc}, accuracy MAE for NYISO increases slightly (10\%) as the number of customers in the treatment group decreases. However, this slight increase can be attributed to the slight increase in the average consumption (+6.5\%) in Table \ref{averageconsumption}. In this study, the statement that the number of customers in the treatment group decreases is equivalent to say that the number of customers in the control group increases, and they can be used Interchangeably.

On the other hand, the accuracy MAE of RCT improves significantly as the number of customers in the control group increases. The accuracy MAE shows 28\% decrease from 5\% to 25\%, which means as the control group is becoming larger, it is able to produce better CBL for the treatment group. According to the results in Fig. \ref{granularbias}, bias for NYISO, also, shows a slight increase. The bias for NYISO is positive, and it stays positive for different groups. On the other hand, bias for RCT changes drastically. As the number of customers in the control group increases, the sign of bias value changes from positive to negative. It is worth mentioning that since all the event days have higher total consumption than their prior days, it is expected that CBL has a negative bias. Therefore, the fact that NYISO shows a positive bias is not a good outcome for this method.   

\begin{table}[]
	\centering
	\caption{ACCURACY “MAE”, BIAS AND OPI OF CLASSIC CBL METHODS AT EVENT HOURS FOR THE GRANULAR CASE}
	\label{errorgranular}
	\begin{tabular}{|c|c|c|c|}
		\hline
		CBL Methods & \begin{tabular}[c]{@{}c@{}}Accuracy MAE\\ (kWh/hour)\end{tabular} & \begin{tabular}[c]{@{}c@{}}Bias\\ (kWh/hour)\end{tabular} & \begin{tabular}[c]{@{}c@{}}OPI\\ (kWh/hour)\end{tabular} \\ \hline
		NYISO5\% & 1.17 & 0.11 & 0.64 \\ \hline
		RCT5\% & 1.58 & 0.22 & 0.90 \\ \hline
		NYISO10\% & 1.19 & 0.12 & 0.66 \\ \hline
		RCT10\% & 1.42 & -0.01 & 0.71 \\ \hline
		NYISO15\% & 1.24 & 0.13 & 0.68 \\ \hline
		RCT15\% & 1.29 & -0.28 & 0.78 \\ \hline
		NYISO20\% & 1.25 & 0.14 & 0.69 \\ \hline
		RCT20\% & 1.22 & -0.21 & 0.71 \\ \hline
		NYISO25\% & 1.29 & 0.15 & 0.72 \\ \hline
		RCT25\% & 1.14 & -0.28 & 0.71 \\ \hline
	\end{tabular}
\end{table}

According to the results in Fig. \ref{granularOPI}, OPI for NYISO retain the same pattern as accuracy MAE and bias. It will increase slightly as the number of customers in the treatment group decreases. On the other hand, OPI for RCT will decrease as the number of customers in the control group increases. Another observation from this figure is that the value of OPI for both NYISO and RCT methods for control groups with 20\% and 25\% are approximately the same. Therefore, as mentioned earlier, because of lower administrative costs, RCT is more preferable. In the next subsection, the performance of RCT and NYISO are explored for the aggregated consumption.  

\subsection{Aggregated Case}
As mentioned earlier, in the aggregated case, the PTR program aggregates the historical consumption data of the customers and perform the CBL calculation and payment settlement in the aggregated level. Therefore, in this case, for all 12 event days, the consumption of all customers in the treatment group are aggregated, then, the CBL is calculated in the aggregated form.

\begin{figure}[b!]
	\centering
	\includegraphics[clip,width=\columnwidth]{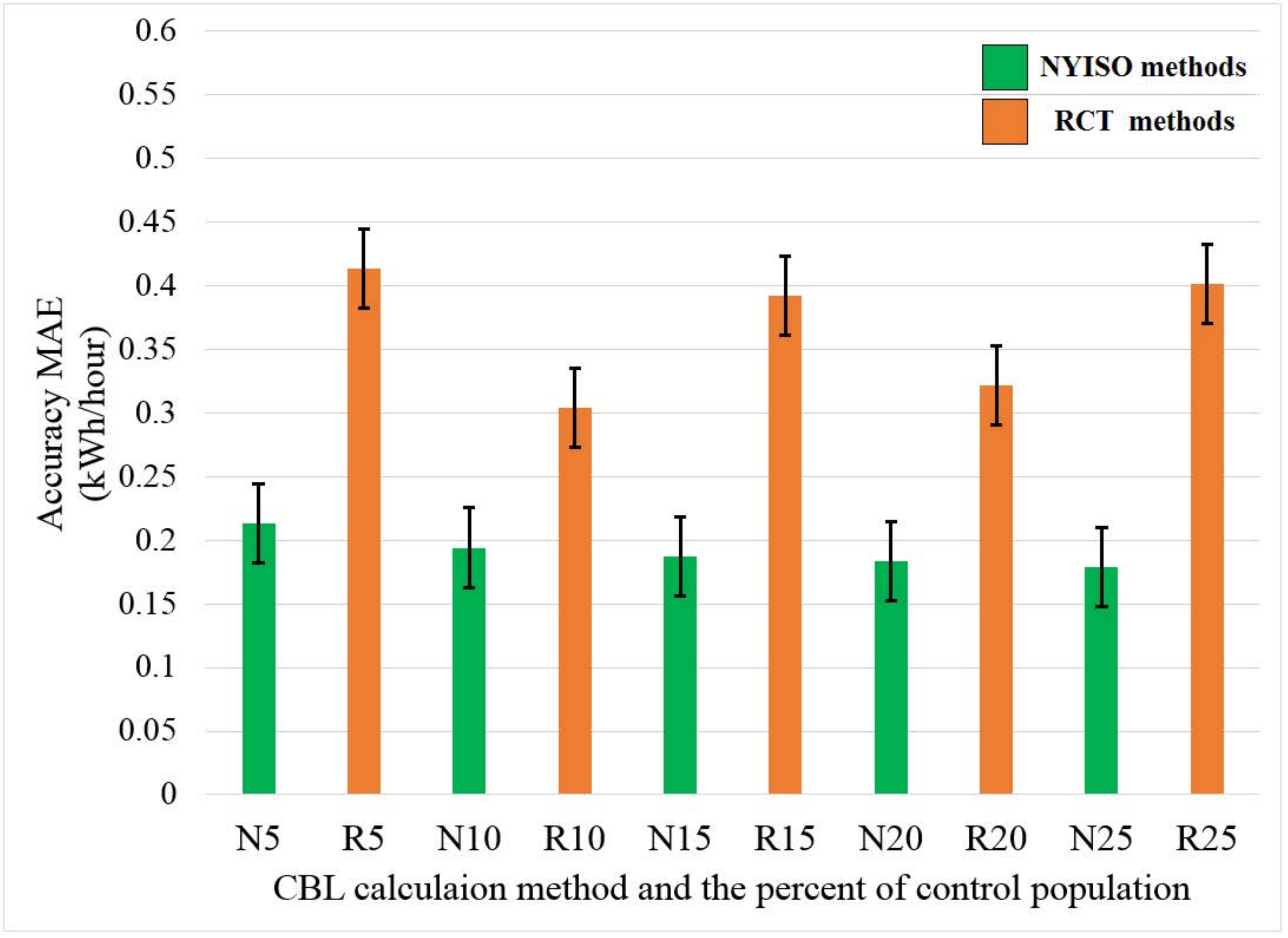}%
	\caption{Accuracy MAE for different CBL calculation with different percent of control population for aggregated case}
	\label{aggregatedacc}
\end{figure}

\begin{figure}[b!]
	\centering
	\includegraphics[clip,width=\columnwidth]{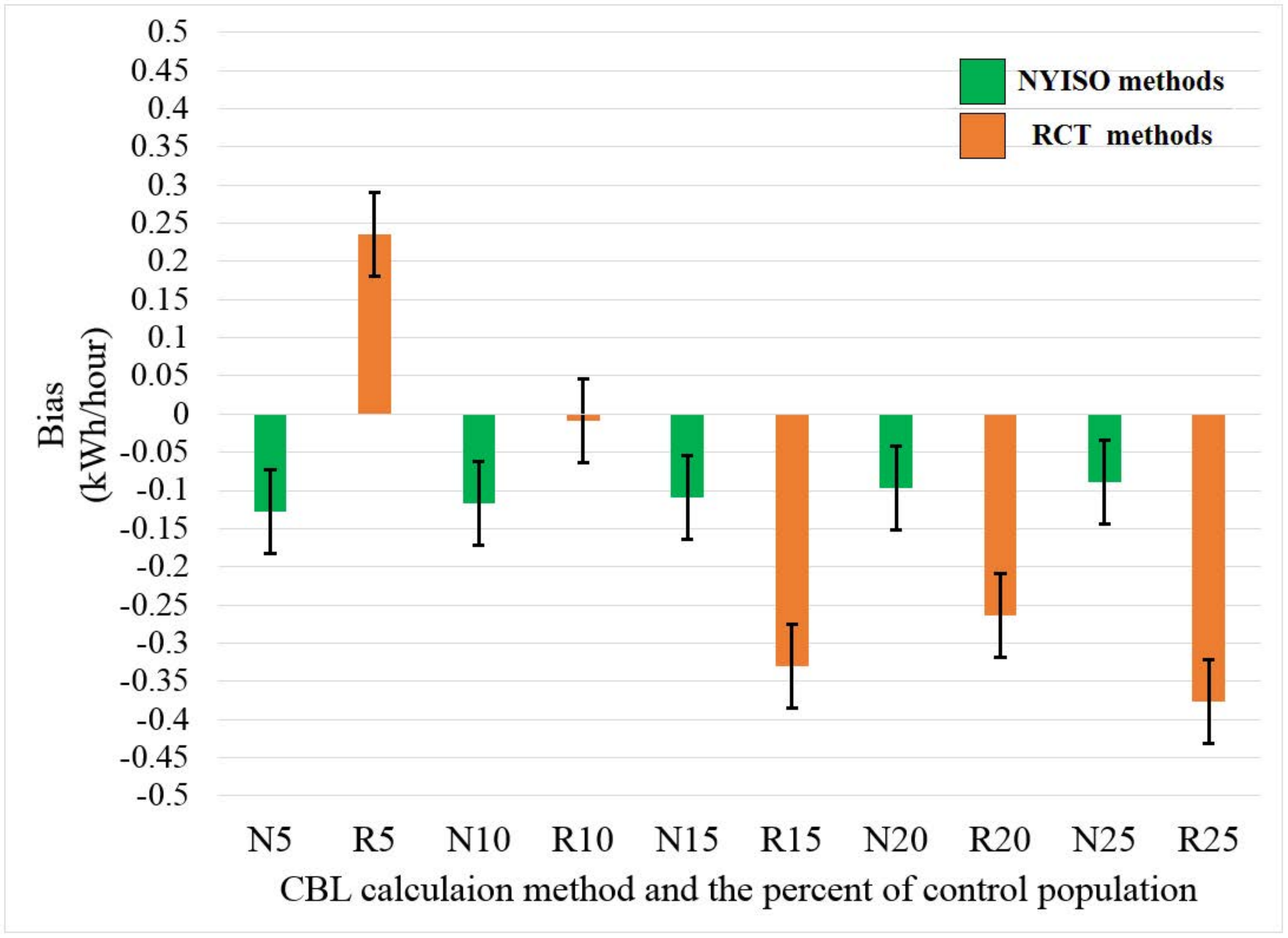}%
	\caption{Bias value for different CBL calculation with different percent of control population for aggregated case}
	\label{aggregatedbias}
\end{figure}

\begin{figure}[t!]
	\centering
	\includegraphics[clip,width=\columnwidth]{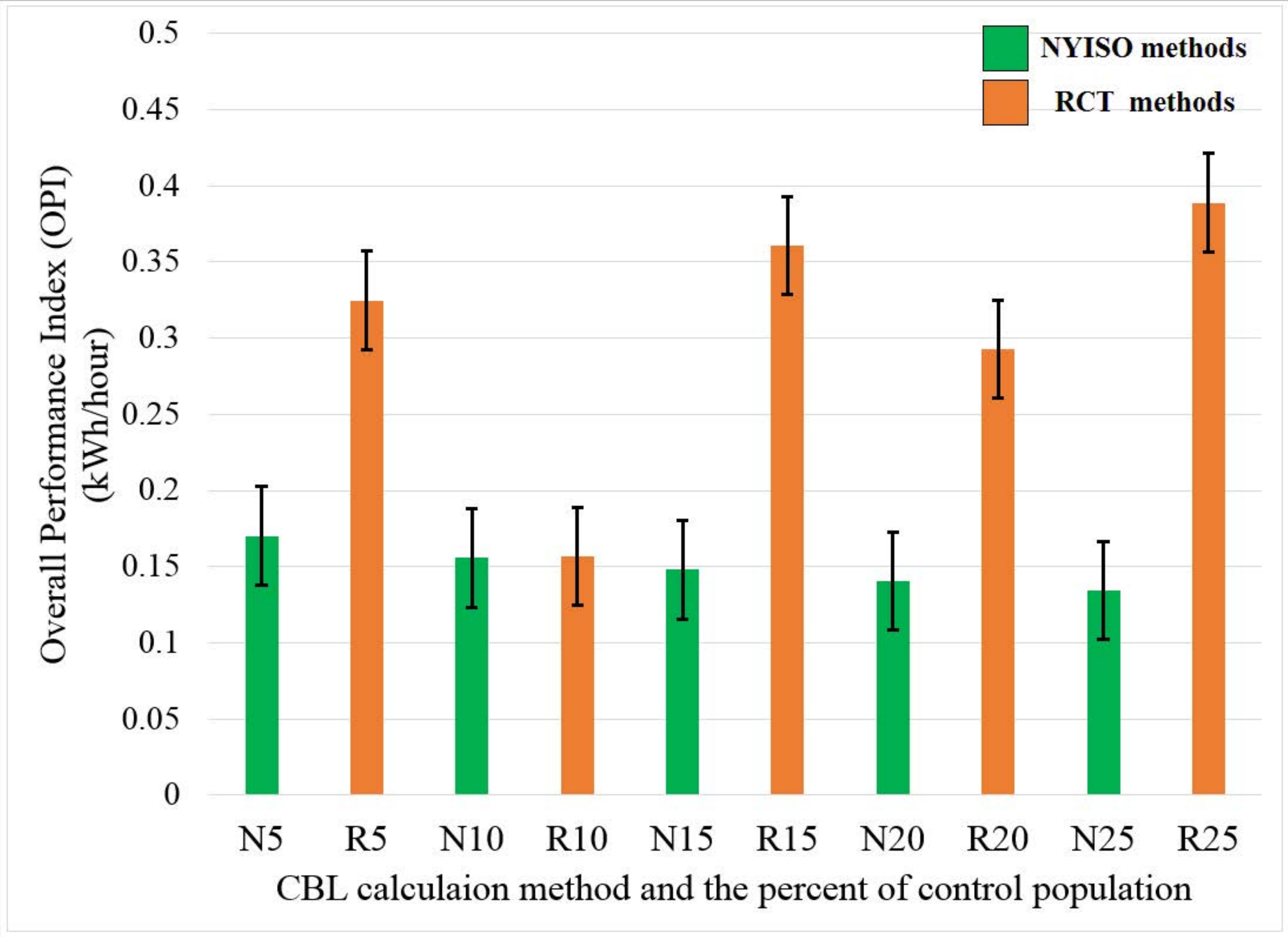}%
	\caption{OPI for different CBL calculation with different percent of control population for aggregated case}
	\label{aggregatedOPI}
\end{figure}

In NYISO, the historical consumption data of customers in the treatment group are aggregated and used as a basis for CBL calculation. In RCT, the aggregated consumption of customers in the control group are used for CBL calculation. Table II lists the value of three error metrics of the CBL calculations for the aggregated case. Similar to the previous case, the information provided in this table is, also, illustrated in Figs. \ref{aggregatedacc}-\ref{aggregatedOPI}.  As mentioned earlier, error bars in each bar plot represents 95\% confidence interval around the adjusted mean for all 12 event days.

\begin{table}[]
	\centering
	\caption{ACCURACY “MAE”, BIAS AND OPI OF CLASSIC CBL METHODS AT EVENT HOURS FOR THE AGGREGATED CASE}
	\label{erroraggregated}
	\begin{tabular}{|c|c|c|c|}
		\hline
		CBL Methods & \begin{tabular}[c]{@{}c@{}}Accuracy MAE\\ (kWh/hour)\end{tabular} & \begin{tabular}[c]{@{}c@{}}Bias\\ (kWh/hour)\end{tabular} & \begin{tabular}[c]{@{}c@{}}OPI\\ (kWh/hour)\end{tabular} \\ \hline
		NYISO5\% & 0.21 & -0.12 & 0.17 \\ \hline
		RCT5\% & 0.41 & 0.23 & 0.32 \\ \hline
		NYISO10\% & 0.19 & -0.11 & 0.15 \\ \hline
		RCT10\% & 0.30 & -0.01 & 0.15 \\ \hline
		NYISO15\% & 0.18 & -0.10 & 0.14 \\ \hline
		RCT15\% & 0.39 & -0.32 & 0.36 \\ \hline
		NYISO20\% & 0.18 & -0.09 & 0.14 \\ \hline
		RCT20\% & 0.32 & -0.26 & 0.29 \\ \hline
		NYISO25\% & 0.17 & -0.08 & 0.13 \\ \hline
		RCT25\% & 0.40 & -0.37 & 0.38 \\ \hline
	\end{tabular}
\end{table}

Fig. \ref{aggregatedacc} shows a major improvement in the accuracy MAE of both methods compared to the previous case. However, the performance of RCT is far cry from the performance of NYISO method. While NYISO shows a consistent 0.2 kWh/hour accuracy MAE for all control populations, RCT shows a variable accuracy MAE of 0.3 to 0.4 for all control populations.  Using the information in Table I, the value of 0.2 kWh/hour accuracy MAE in NYISO is almost equal to 11\% accuracy. In other words, this method is able to capture load changes above 11\%. 

Fig. \ref{aggregatedbias} shows that the bias for NYISO is negative, and it is almost -0.1 kWh/hour. On the other hand, similar to the granular case, the bias for RCT changes drastically and its sign changes from positive to negative as the number of customers in the control group increases.

According to Fig. \ref{aggregatedOPI}, the OPI for NYISO is almost unchanging. On the other hand, the OPI for the RCT changes randomly. Therefore, compared to NYISO, this method is less reliable for the aggregated case, and the NYISO outperforms the RCT significantly in the aggregated case. 

\section{Case study}
In order to investigate the impact of the CBL's error on the financial performance of a DR program, in this paper, a case of hypothetical PTR program is selected. PTR program, as discussed earlier, focuses on load reduction and rewards it. The customers are allowed to consume more than their CBL, and they would be charged according to the fixed tariff. In this paper, it is assumed that the PTR program pays \$0.35/kWh as an incentive for any kWh load reduction and charges the fixed tariff of \$0.097/kWh for electricity consumption. These values are employed by a real PTR program in Anaheim Public Utility (APU) pilot project for the residential customers \cite{faruqui}. The hypothetical PTR program, used in this paper, is assumed to be offered to the dataset described earlier. 

\section{Financial analysis}
In this section, the financial performance of PTR with two CBL calculation methods for two cases of granular and aggregated loads are assessed. 
\subsection{Granular Case}
In the granular case, as mentioned earlier, the PTR program calculates the CBL for each individual customer. Therefore, in this case, for all 12 event days, the CBL for all customers in different datasets is calculated. 

The customers in DR programs are anticipated to respond to the incentive and reduce their demand. If the CBL is calculated accurately (i.e. MAE=0), the load reduction can be attributed to the PTR incentive effect. However, if the CBL is inaccurate, the load reduction is comprised of two components; one component is in response to the incentive effect and the other is due to the CBL inaccuracy. In this paper, because the focus is on the accuracy of CBL calculation methods, a dataset without DR event is selected. Therefore, it is possible to claim that the first component is zero; hence, the perceived load reduction is all because of the second component (i.e. CBL inaccuracy) which is defined as false load reduction. In this section, false load reduction represents the occasions that the CBL is higher than the load. the other occasions when the load is higher than the CBL are not considered in the study because PTR program does not pay any money for those moments. Therefore, they are deliberately left out in the financial analysis. It is worth mentioning that the impact of those time intervals when the load is higher than CBL is reflected in the error analysis. As discussed earlier, the aforementioned load reductions are called "false" because the actual load reduction is zero.

\begin{table}[b]
	\centering
	\caption{LOAD REDUCTION AND “PTR” PAYMENT SETTLEMENT AS A PERCENT OF TOTAL CONSUMPTION AND TOTAL REVENUE FOR GRANULAR CASE}
	\label{financialgranular}
	\begin{tabular}{|c|c|c|}
		\hline
		CBL Methods & \begin{tabular}[c]{@{}c@{}}False Load Reduction\\ as a Percentage of \\ Consumption on \\ Event Day (\%)\end{tabular} & \begin{tabular}[c]{@{}c@{}}Rebate as a Percentage\\  of Utility Revenue (\%)\end{tabular} \\ \hline
		NYISO5\% & 31.99 & 115.44 \\ \hline
		RCT5\% & 42.70 & 154.07 \\ \hline
		NYISO10\% & 32.03 & 115.60 \\ \hline
		RCT10\% & 35.16 & 126.86 \\ \hline
		NYISO15\% & 31.65 & 114.21 \\ \hline
		RCT15\% & 27.01 & 97.48 \\ \hline
		NYISO20\% & 31.27 & 112.86 \\ \hline
		RCT20\% & 27.47 & 99.14 \\ \hline
		NYISO25\% & 30.95 & 111.69 \\ \hline
		RCT25\% & 24.42 & 88.13 \\ \hline
	\end{tabular}
\end{table}

 \begin{figure}[b!]
 	\centering
 	\includegraphics[clip,width=\columnwidth]{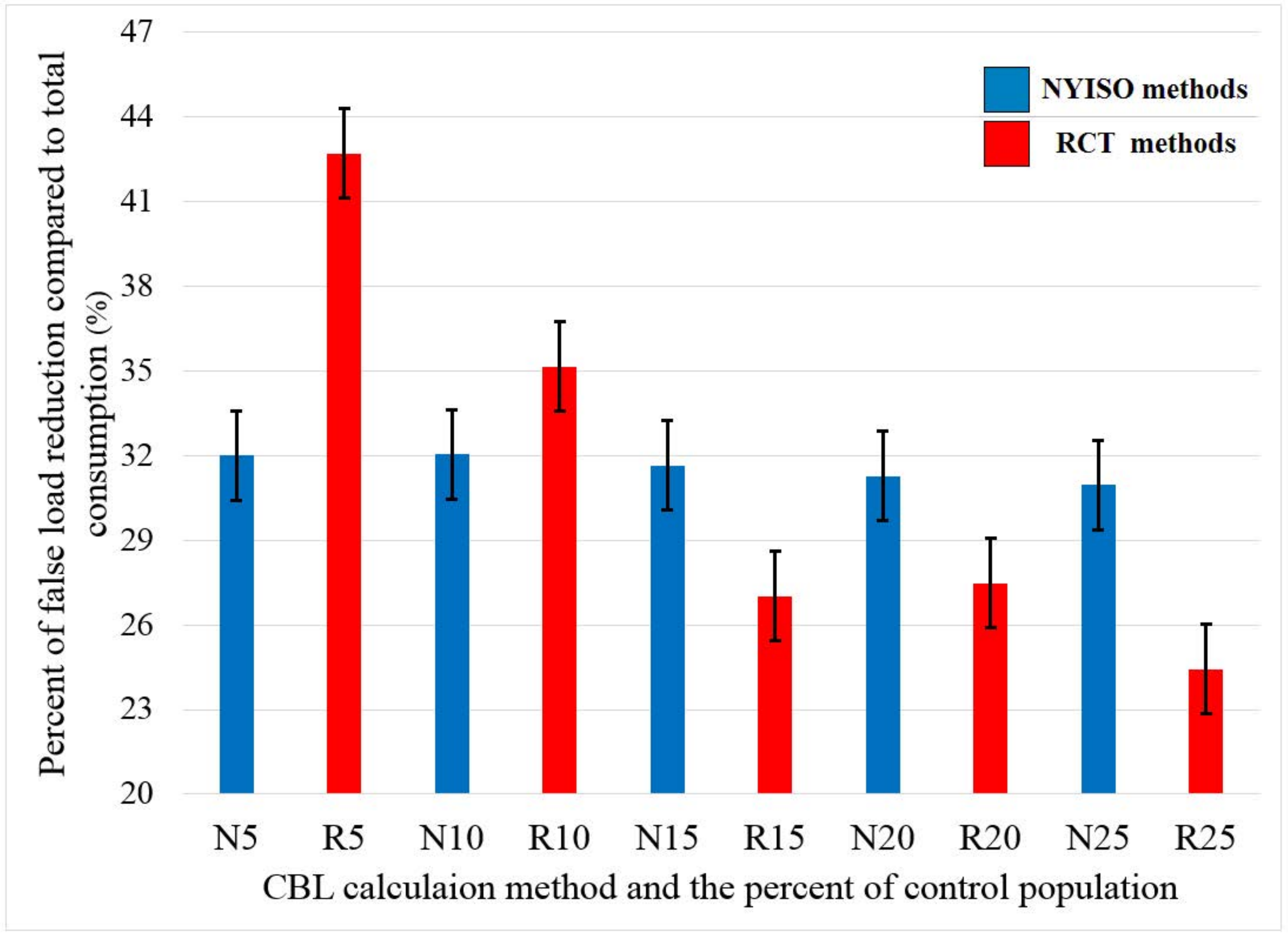}%
 	\caption{Percent of false load reduction compared to total consumption for different CBL calculation with different percent of control population for granular case}
 	\label{granularFLR}
 \end{figure}

 \begin{figure}[h!]
 	\centering
 	\includegraphics[clip,width=\columnwidth]{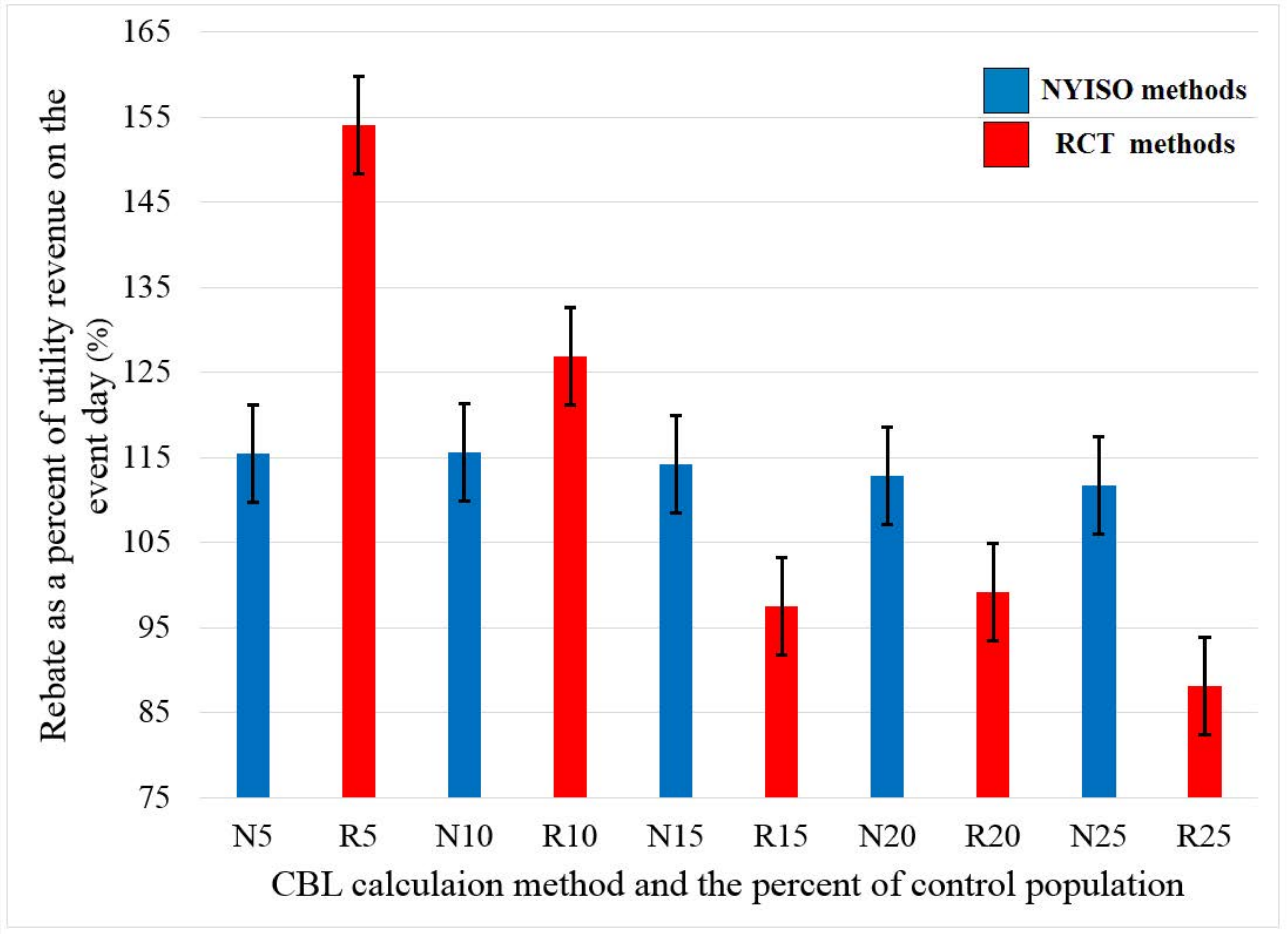}%
 	\caption{Rebate as percent of utility revenue for different CBL calculation with different percent of control population for granular case}
 	\label{granularRR}
 \end{figure}
 
Table \ref{financialgranular} lists the false load reductions as a percentage of total consumption on the event day under NYISO and RCT methods for different datasets for the granular case. It, also, shows how much rebate, as a percentage of utility revenue, this utility must pay to these customers on the event day.

For the illustration purpose, the results of this table are shown in Fig. \ref{granularFLR}-\ref{granularRR}. Whiskers in each bar plot represents 95\% confidence interval around the adjusted mean for all 12 event days. As discussed earlier, all the rebate money is incurred because of CBL inaccuracy. 

According to the results of Fig. \ref{granularFLR}, the percent of false load reduction compared to total consumption for NYISO method for different percent of control population is almost constant (32\%). On the other hand, this value for RCT method decreases significantly as the number of customers in the control group increases (42\% decrease). It is worth mentioning that the false load reduction is directly correlate to the bias value. As discussed earlier, negative bias value indicates that the method under-estimates the CBL. Holding all other independent variables constant, The lower CBL means the lower load reduction. Therefore, the last observation about RCT could be attributed to the decreasing trend of bias values in Fig. \ref{granularbias}. 

According to the results of Fig. \ref{granularRR}, the rebate as a percentage of utility revenue for NYISO method for different percent of control population is almost constant (115\%). On the other hand, this value, similar to the percent of false load reduction compared to total consumption, for RCT method decreases significantly as the number of customers in the control group increases (42\% decrease), which,similarly, could be attributed to the decreasing trend of bias values in Fig. \ref{granularbias}.
 
\subsection{Aggregated Case}
\begin{figure}[b!]
 	\centering
 	\includegraphics[clip,width=\columnwidth]{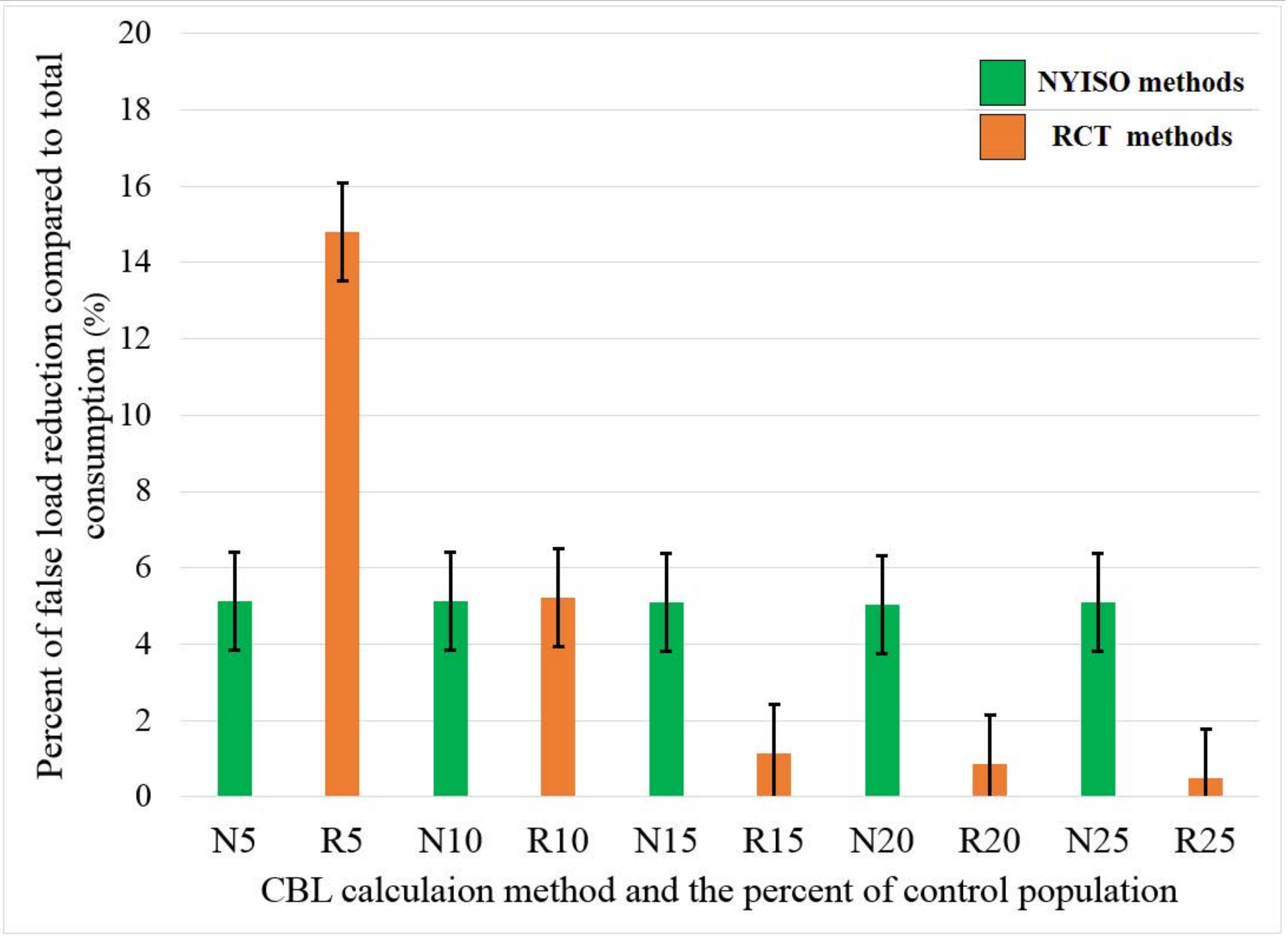}%
 	\caption{Percent of false load reduction compared to total consumption for different CBL calculation with different percent of control population for aggregated case}
 	\label{aggregatedFLR}
 \end{figure}

\begin{figure}[t!]
 	\centering
 	\includegraphics[clip,width=\columnwidth]{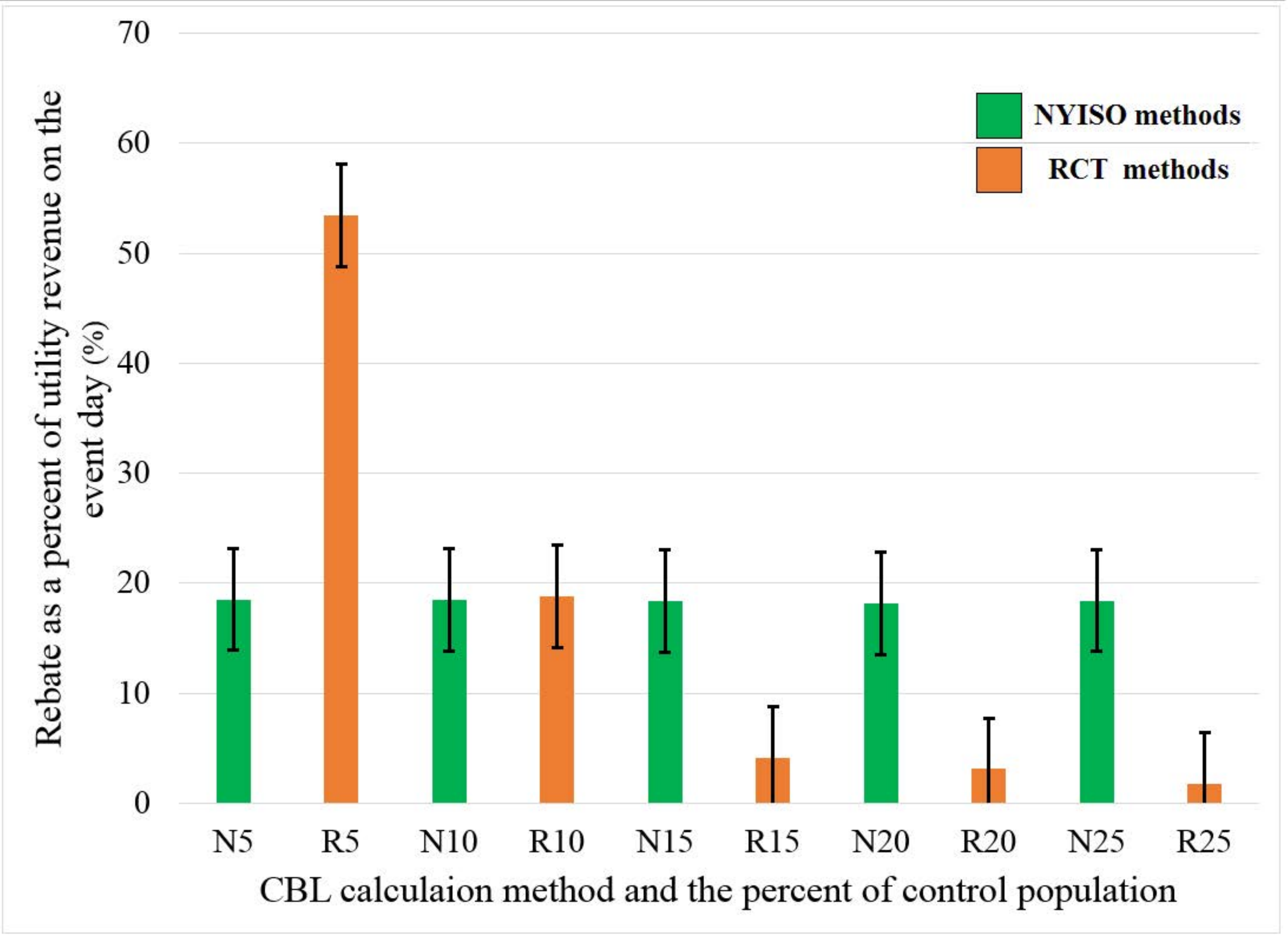}%
 	\caption{Rebate as percent of utility revenue for different CBL calculation with different percent of control population for aggregated case}
 	\label{aggregatedRR}
 \end{figure}

In the aggregated case, as mentioned earlier, the PTR program performs the CBL calculation after aggregating the historical consumption data of the customers. In this case, for all 12 event days, the CBL is calculated in the aggregated form. Table \ref{financialaggregated} lists the false load reductions as a percentage of total consumption on the event day of the NYISO and RCT methods for different datasets. It, also, shows how much rebate, as a percentage of utility revenue, this utility must pay to these customers on the event day. The information in this table is, also, illustrated in Figs. \ref{aggregatedFLR}-\ref{aggregatedRR}.   Whiskers in each bar plot represents 95\% confidence interval around the adjusted mean for all 12 event days.

\begin{table}[b]
	\centering
	\caption{LOAD REDUCTION AND “PTR” PAYMENT SETTLEMENT AS A PERCENT OF TOTAL CONSUMPTION AND TOTAL REVENUE FOR AGGREGATED CASE}
	\label{financialaggregated}
	\begin{tabular}{|c|c|c|}
		\hline
		CBL Methods & \begin{tabular}[c]{@{}c@{}}False Load Reduction\\ as a Percentage of \\ Consumption on \\ Event Day (\%)\end{tabular} & \begin{tabular}[c]{@{}c@{}}Rebate as a Percentage\\  of Utility Revenue (\%)\end{tabular} \\ \hline
		NYISO5\% & 5.13 & 18.53 \\ \hline
		RCT5\% & 14.81 & 53.45 \\ \hline
		NYISO10\% & 5.12 & 18.47 \\ \hline
		RCT10\% & 5.20 & 18.79 \\ \hline
		NYISO15\% & 5.09 & 18.37 \\ \hline
		RCT15\% & 1.14 & 4.11 \\ \hline
		NYISO20\% & 5.03 & 18.15 \\ \hline
		RCT20\% & 0.85 & 3.09 \\ \hline
		NYISO25\% & 5.10 & 18.42 \\ \hline
		RCT25\% & 0.49 & 1.80 \\ \hline
	\end{tabular}
\end{table}

According to Fig. \ref{aggregatedFLR}, the percent of false load reduction compared to total consumption for NYISO method is 5\%, which shows a significant change (almost 6 times) compared to the granular case. Since, for the aggregated case, NYISO has given a negative bias, this significant change was expected.  However, it is worth reminding that the percent of false load reduction is not a metric for evaluating the performance of a CBL calculation method in general, and any conclusions from this figure about the performance of the CBL calculation is only applicable to the context of this particular hypothetical PTR program.

As mentioned in the error analysis of RCT in the aggregated case, the values for RCT seems to change randomly. The variable nature of the values in RCT method shows that this method is less reliable in aggregated cases. Because of the large negative bias, it is expected that the percent of false load reduction compared to total consumption be very small and close to zero. Fig. \ref{aggregatedRR} is the monetary translation of Fig. \ref{aggregatedFLR}, and the rebate as a percentage of utility revenue for NYISO method for different control populations is almost 18\%.
\section{conclusion}
FERC Order 745 allows demand response owners to sell their load reduction in the wholesale market. However, in order to be able to sell the load reduction, many implementation challenges must be addressed, one of which is to establish CBL calculation methods with acceptable error performance, which has proven to be very challenging so far. In this paper, the error and financial performance of Randomized Controlled Trial (RCT) method, applied to both granular and aggregated forms of the consumption load, are investigated for a hypothetical demand response program offered to a real dataset of residential customers. These customers, due to non-correlated personal and household activities, show different characteristics compared to the industrial and commercial customers; therefore, it is critical to study these customers separately. 

For the purpose of analysis and comparison, one well-established CBL calculation method of HighXofY (NYISO) is selected to provide a basis for comparison with RCT and. Then,  by using a real dataset of residential customers' consumption data, the empirical error and financial analyses are carried out.
 
The key conclusions of this paper are:

\begin{itemize}
\item The RCT method with larger control population has much better error performance compared to the smaller control population in the granular case; 
\item The RCT method is almost insensitive to the control population in the aggregated case;
\item In the granular case, OPI for NYISO and RCT methods for control groups of 20\% and 25\% are approximately the same. Therefore, if error performance is the major concern, because of lower administrative costs, RCT is more preferable CBL calculation method than NYISO;
\item the NYISO method shows a better and more consistent error performance compared to the RCT for the aggregated case; 
\item The error and financial performance of both methods in the aggregated case are significantly better than the granular case. Therefore, for applications and studies that do not need an individual payment settlement, it is much better to perform the analysis on the aggregated level.
\end{itemize}

In future work, the authors plan to assigns treatment and control groups' members into different clusters and use data analytics to match clusters within control group with clusters within treatment group. Methods such as Discrete Wavelet Transform (DWT) are able to treat consumption data as signals and extract features from these signals. These features could be utilized for the classification of the members in treatment and control groups.  Matching clusters with the similar features is possible to produce better results. Also, it is important to study the performance of CBL calculation methods in the presence of a real DR event. Finding the CBL accurately in the presence of a real DR program is another major concern in EM\&V of DR programs. Moreover, developing techniques to harness the randomness of the residential customers' data in the granular form is another interesting direction, which authors plan to explore.

\ifCLASSOPTIONcaptionsoff
  \newpage
\fi

\bibliographystyle{IEEEtran}
\bibliography{Bi}


\end{document}